\documentclass{aa}
\usepackage[dvips]{graphicx}
\begin{document}
   \title{CCD $uvby\beta$ photometry of young open clusters}

   \subtitle{I.The double cluster $h$ and $\chi$ Persei
    \thanks{Tables 6 and 7 will be only available in electronic form at
the CDS. Now are available at
http://pleione.uv.es/recent.html}}

   \author{G.Capilla\inst{1} \and J.Fabregat\inst{1,2}}

   \offprints{J. Fabregat} 
   \mail{juan@pleione.uv.es}

   \institute{Departamento de Astronom\'\i a,
              Universidad de Valencia, 46100 Burjassot,
       Valencia, Spain\\
    \and
           GEPI/FRE K2459 du CNRS, Observatoire de Paris-Meudon, 92195
      Meudon Principal Cedex, France}

   \date{Received date; accepted date}

       \maketitle

   \begin{abstract}
   We present CCD $uvby\beta$ photometry for stars in the nuclei of the
   young double cluster $h$ and $\chi$ Persei. We find that the reddening
   is highly variable through the $h$ Per nucleus, increasing from west to
   east, with values ranging from $E(b-y) = 0.328\pm0.022$ in the
   western part to $E(b-y) = 0.465\pm0.024$ in the south-east. Towards  
   $\chi$ Per the reddening is fairly constant, with $E(b-y) = 
   0.398\pm0.025$. Both clusters share a common distance modulus of
   11.7$\pm$0.1 mag., and an age of $\log t = 7.10\pm0.05$ years.

   \end{abstract} 

%
%________________________________________________________________

    \section{Introduction}

The precise determination of galactic open clusters main physical
parameters plays a central role in the study of the stellar structure and
evolution. With accurate photometric data, and once the external variables
such as reddening are corrected for, the cluster distances, ages and
chemical abundances can be inferred from the study of the photometric
colour-magnitude and colour-colour diagrams. 

The usual way to obtain the cluster age is by means of isochrone fitting
to the main sequence in a colour-magnitude diagram. In the case of
young clusters -younger than 50 Myr- isochrone fitting is made difficult by
the usual presence of differential reddening across the cluster face,
which widens the observed main sequence. Moreover, the presence of
emission line stars, like Be o PMS stars, which occupy anomalous positions
in the photometric diagrams, additionally contributes to a further main
sequence widening. Hence, the fit of a particular isochrone can be a very
uncertain process, and it is not difficult to find recent age
determinations with very diverging values for a given young cluster.

In a recent paper, Fabregat \& Torrej\'on (\cite{fabregat00}) propose the
use of isochrone fitting in the $V_0 - c_0$ plane of the $uvby$
photometric system as an adequate tool to obtain accurate cluster ages. 
The range of variation of the $c_0$ index along the B-type sequence
amounts more than 1 mag., being significantly larger than most common used
photometric colours. Moreover, the $c_0$ index is less affected by
reddening, and allows an efficient segregation of emission-line stars. 

In order to produce accurate and homogeneous dating for a sample of young
galactic clusters, we have undertaken an observational programme to
obtain CCD $uvby\beta$ photometry. In this paper we present the first
results, related to the clusters $h$ and $\chi$ Persei.

The double cluster \object{$h$} and \object{$\chi$ Persei} is one of the
richest young open clusters in the Galaxy, and also one of the brightest
and closest to us. On a clear night, far from the light pollution, it can
be easily seen with a naked eye, distinctly shining in the Milky Way
between Perseus and Cassiopeia constellations. $h$ and $\chi$ Persei
(\object{NGC 869} and \object{NGC 884} respectively) form the nucleus of
the broader Per OB1 association. 

The published work on $h$ and $\chi$ Persei is very extensive. A summary
of the key papers and discussion on the past work is given by Waelkens et
al. (\cite{waelkens}). Within the literature there is not convergence on
the fundamental parameters of the clusters, such as their distances and
ages. The discrepancies in the cluster distance moduli are well in excess
of 0.5 mag., while there is not agreement about both clusters being of the
same or different ages.

Two modern studies based on CCD photometry, presented by Keller et
al. (\cite{keller}) and  Marco \& Bernabeu (\cite{marco}), converge in a
common distance modulus of about 11.7 mag. for both clusters, but still
present controversial results regarding their ages. The former finds a
common age of log $t$ = 7.1 years for both clusters and the surrounding
population, while the latter claims the existence of at least three
different episodes of star formation. To contribute to ascertain this
issue is one of the objectives of the present paper.

%__________________________________________________________________

\section{Observations and reduction procedure}
%_________________________________________________________________________
\subsection{Observations and image processing}

CCD photometry of the central regions of $h$ and $\chi$ Persei was
obtained on the nights 20 to 22 November 1998 at the 1.52m. telescope of
the Observatorio Astron\'omico Nacional, located at the Calar Alto
Observatory (Almer\'{\i}a, Spain). The chip employed was the Tektronics
TK 1024 AB, with a size of 1024\,x\,1024 pixels. The 0\farcs\,4 unbinned
pixels provide a field size of 6\farcm\,9x6\farcm\,9, which almost
entirely covers the clusters 'nuclei' area, as defined by Maps 2 and 4 
in Oosterhoff (\cite{oost}). 

Observations were done through the four Str\"omgren $uvby$ and
Crawford narrow and wide H$\beta$ filters, being every field
sequentially measured through the six filters. Two different
exposure times were used with each filter, in order to ensure a wide
range of stellar magnitudes. Exposure times in each filter were
selected so as a B type star produces approximately equal count rates
through all filters. Employed exposure times are presented in Table
\ref{t1}.

%______________________________________________________________________
\begin{table}
\centering
\caption[]{Exposure times with the different filters used, in seconds.}
\label{t1}
\begin{tabular}{lrr}
\hline 
\noalign{\smallskip}
Filter & short & large \\
\noalign{\smallskip}
\hline
\noalign{\smallskip}
$y$ & 10 & 50  \\
$b$ & 12 & 60  \\
$v$ & 35 & 175 \\
$u$ & 120 & 600  \\
H$\beta_{w}$ & 12 & 60 \\
H$\beta_{n}$ & 30 & 150  \\
\noalign{\smallskip}
\hline 
\end{tabular}
\end{table}
%_____________________________________________________________________

In order to ensure the atmospheric extinction and standard transformation
determination, three additional fields centered on the open clusters
\object{NGC 1039}, \object{NGC 6910} and \object{NGC 6913} were also
observed at different airmasses. The list of all observations is presented
in Table \ref{t2}.

%________________________________________________________________________
\begin{table}
\centering
\caption[]{List of all observed fields}
\label{t2}
\begin{tabular}{lcccl}
\hline 
\noalign{\smallskip}
NGC & JD & date & airmass & int. time \\ 
\noalign{\smallskip}
\hline
\noalign{\smallskip}
869 & 51138 & 20-11-98 & 1.06 & short\\
869 & 51138 & 20-11-98 & 1.07 & long\\
869 & 51138 & 20-11-98 & 1.83 & short\\
869 & 51139 & 21-11-98 & 1.07 & short\\
869 & 51139 & 21-11-98 & 1.07 & long\\
869 & 51140 & 22-11-98 & 1.07 & short\\
869 & 51140 & 22-11-98 & 1.63 & short\\
\noalign{\smallskip}
884 & 51138 & 20-11-98 & 1.10 & short\\
884 & 51138 & 20-11-98 & 1.12 & long\\
884 & 51138 & 20-11-98 & 1.98 & short\\
884 & 51139 & 21-11-98 & 1.11 & short\\
884 & 51139 & 21-11-98 & 1.13 & long\\
884 & 51140 & 22-11-98 & 1.06 & short\\
884 & 51140 & 22-11-98 & 1.70 & short\\
\noalign{\smallskip}
1039 & 51138 & 20-11-98 & 1.57 & short\\
1039 & 51140 & 22-11-98 & 1.55 & short\\
\noalign{\smallskip}
6910 & 51138 & 20-11-98 & 1.11 & short\\
6910 & 51138 & 20-11-98 & 1.77 & short\\
6910 & 51138 & 20-11-98 & 1.14 & long\\
6910 & 51138 & 20-11-98 & 1.83 & long\\
6910 & 51140 & 22-11-98 & 1.14 & short\\
6910 & 51140 & 22-11-98 & 1.66 & short\\
\noalign{\smallskip}
6913 & 51138 & 20-11-98 & 1.25 & short\\
6913 & 51138 & 20-11-98 & 1.30 & long\\
6913 & 51140 & 22-11-98 & 1.18 & short\\
6913 & 51140 & 22-11-98 & 1.81 & short\\
\noalign{\smallskip}
\hline
\end{tabular}
\end{table}

%_________________________________________________________________________

Images were processed using IRAF.\footnote{IRAF is distributed by the
National Optical Astronomy Observatories, which are operated by the
Association of Universities for Research in Astronomy, Inc., under
cooperative agreement with the National Science Foundation, U.S.A.}
A sizeable sample of bias and sky flat frames were obtained at the
beginning and at the end of every night. The images were subjected to the
usual overscan, bias and flat field corrections. 

Photometry was performed using the DAOPHOT package (Stetson
\cite{stetson}). Aperture photometry was obtained for a number of
sufficiently clean stars in each frame, through a constant 14 pixel
radius which was chosen to contain virtually all the stellar flux in all
images, as indicated by a grow curve analysis. PSF-fitting photometry was
subsequently obtained for all identified stars in all frames. A constant
PSF provided a good representation of the stellar profiles through each
frame. The difference between aperture and PSF-based instrumental
magnitudes were determined for each frame, and the latter were corrected
accordingly.  

\subsection{Extinction and instrumental system}

The atmospheric extinction was determined by the multi-night, multi-star
method described by Manfroid (\cite{manf93}). Computations were done by
using the RANBO2 package, written by J. Manfroid.  The implementation of
this reduction procedure allows the construction of a consistent natural
system, which contains the extra-atmospheric instrumental magnitudes of all
constant stars included in the computation procedure. Stars from all
observed fields were included in the building of the natural system.

The value of the extinction coefficient was determined for each individual
frame. For the stars in common, the mean difference between the observed
and natural magnitudes was obtained, and divided by the airmass to obtain 
the corresponding extinction coefficient.

%___________________________________________
\begin{table*}
\centering
\caption[]{List of standard stars observed and transformed to the $uvby$
and H$\beta$ standard systems. Columns 7 to 11 give the transformation
residuals, in the sense standard value minus transformed one. N is the
number of measures of each standard.} 
\label{t3}
\begin{tabular}{crrrrrrrrrrr}
\hline 
\noalign{\smallskip}
& & & & & & & & $D$ & & &  \\
star & $V$ & $(b-y)$ & $m_{1}$ & $c_{1}$ & $\beta$ & $V$ & $(b-y)$ &
$m_{1}$ & $c_{1}$ & $\beta$ & N  \\
\noalign{\smallskip}
\hline
\noalign{\smallskip}
0869-0837 & 14.090 &   -     &  -      &  -    & -     & 
-0.010 &  -   &  -   &  -   &   -  & 4      \\
0869-0843 &  9.317 &  0.286 & -0.065 &  0.172 & 2.593  & 
 0.003 & -0.009 &  0.015 & -0.006 & -0.026 & 2     \\
0869-0922 &  -      &  0.304 & -0.066 &  0.130 & -     & 
 -    &  0.006 & -0.016 &  0.001 &  -   & 3      \\
0869-0963 &  -      &  0.286 & -0.044 &  0.191 & -     &  
-    & -0.003 & -0.010 & -0.003 &  -   & 5       \\
0869-0978 &  -      &  0.301 & -0.033 &  0.167 & 2.627 &  
-    &  0.004 & -0.006 &  0.010 & -0.016 & 4     \\
0869-0980 &  -      &  0.284 & -0.042 &  0.178 & -     & 
 -    &  0.006 & -0.012 & -0.011 &  -   & 3      \\
0869-0991 &  -      &  0.328 & -0.055 &  0.255 & -     &
 -    &  0.001 & -0.007 &  0.020 &  -   & 5       \\
0869-1004 &  -      &  0.305 & -0.051 &  0.204 & -     & 
 -    &  0.013 & -0.010 &  0.010 &  -   & 5       \\
0869-1181 &  -      &  0.350 & -0.049 &  0.374 & 2.730 &  
-    &  0.022 &  0.015 &  0.005 &  0.012 & 5     \\
0869-1187 & 10.857 &  0.334 & -0.046 &  0.205 & 2.618  & 
-0.037 &  0.014 & -0.017 &  0.007 & -0.030 & 5  \\
\noalign{\smallskip}
0884-2167 & 13.401 &  0.346 & -0.057 &  0.604 & 2.762  & 
-0.041 &  0.006 &  0.001 &  0.023 &  0.010 & 4  \\
0884-2196 & 11.538 &   -     &  -      &  -   & 2.666  & 
 0.032 &  -   &  -   &  -   &  0.004 & 5     \\
0884-2200 &  -      &  -      & -       & -   & 2.721  & 
 -    & -   &  -  &  -   &  0.016 & 5        \\
0884-2232 & 11.102 &   -     &  -      &  -   & 2.639  & 
 0.008 &  -   &  -   &  -   & -0.012 & 4     \\
0884-2235 &  9.361 &  0.330 & -0.095 &  0.134 & 2.608  & 
-0.001 & -0.014 &  0.007 &  0.016 &  0.003 & 3  \\
0884-2246 &  9.930 &  0.291 & -0.058 &  0.124 & 2.625  & 
-0.030 & -0.004 & -0.004 &  0.005 &  0.002 & 4  \\
0884-2251 & 11.560 &  0.315 & -0.051 &  0.371 & 2.708  & 
 0.000 & -0.013 &  0.009 & -0.022 &  0.001 & 5   \\
0884-2296 &  8.499 &  0.323 & -0.104 &  0.136 &  -     & 
 0.031 & -0.032 &  0.032 & -0.003 &   -  & 1     \\
0884-2311 &  9.363 &  0.299 & -0.074 &  0.146 & 2.601  & 
 0.017 & -0.017 &  0.010 &  0.014 & -0.020 & 3   \\
0884-2330 & 11.446 &  0.273 & -0.070 &  0.268 & 2.630  & 
-0.026 & -0.006 &  0.020 &  0.009 &  0.010 & 5   \\
\noalign{\smallskip}
1039-0226 & 10.482 &   -     &  -      &  -   &  -     & 
-0.002 &  -   &  -   &  -   &   -  & 2        \\
1039-0267 & 11.940 &  0.296 &  0.133 &  0.505 & 2.683  &  
0.020 &  0.007 &  0.005 & -0.024 &  0.005 & 2    \\
1039-0274 &  9.745 &  0.092 &  0.159 &  0.994 & 2.876  & 
-0.005 & -0.006 &  0.017 & -0.021 & -0.014 & 2   \\
1039-0278 & 11.789 &  0.146 &  0.202 &  0.911 & 2.857  & 
 0.031 & -0.002 & -0.004 & -0.011 &  0.002 & 2    \\
1039-0284 & 10.741 &  0.151 &  0.185 &  0.898 & 2.862  & 
 0.019 &  0.000 & 0.009 & -0.004 &  0.014 & 2    \\
1039-0294 & 11.185 &  0.186 &  0.195 &  0.782 &  -     & 
 0.035 & -0.010 &  0.009 &  0.014 &   -  & 2     \\
1039-0301 & 10.041 &  0.040 &  0.185 &  1.014 & 2.887  & 
-0.011 &  0.026 & -0.033 & -0.001 & -0.029 & 2   \\
1039-0303 &  9.951 &  0.055 &  0.169 &  1.023 &  -     & 
-0.001 &  0.000 & -0.006 & -0.002 &   -  & 2     \\
\noalign{\smallskip}
6910-0010 & 10.782 &   -     &  -      &  -   & 2.941  & 
-0.032 &  -   &  -   &  -   &  0.032 & 5     \\
6910-0021 & 11.763 &  0.582 & -0.099 &  0.234 & 2.678  & 
-0.033 &  0.008 & -0.001 & -0.014 &  0.026 & 5    \\
6910-0024 & 11.710 &   -     &  -      &  -   & 2.637  & 
 0.010 &  -   &  -   &  -   & -0.010 & 5      \\
6910-0028 & 12.243 &   -     &  -      &  -   &  -     & 
-0.023 &  -   &  -   &  -   &   -  & 5        \\
6910-0041 & 12.803 &  0.759 & -0.166 &  0.319 & 2.674  & 
 0.007 & -0.009 &  0.026 & -0.029 &  0.006 & 5    \\
\noalign{\smallskip}
6913-0001 &  8.842 &  0.726 & -0.159 &  0.166 & 2.596  & 
 0.018 &  0.004 & -0.011 &  0.004 &  0.007 & 3    \\
6913-0002 &  8.912 &  0.644 & -0.148 &  0.112 & 2.625  & 
-0.002 & -0.014 &  0.008 &  0.008 &  0.029 & 3    \\
6913-0003 &  8.942 &  0.699 & -0.186 &  0.165 & 2.639  & 
 0.038 & -0.019 &  0.026 & -0.035 &  0.045 & 3   \\
6913-0004 & 10.199 &  0.621 & -0.115 &  0.157 & 2.632  & 
-0.019 & -0.001 & -0.005 &  0.013 &  0.015 & 4     \\
6913-0008 & 12.190 &  0.546 & -0.062 &  0.365 & 2.618  & 
-0.020 &  0.024 & -0.028 & -0.025 & -0.044 & 4    \\
6913-0009 & 11.747 &  0.587 & -0.069 &  0.297 &  -     & 
 0.033 &  0.033 & -0.041 &  0.013 &   -  & 4      \\
6913-0010 &  -      &  -      & -       & -   & 2.678  &  
-    & -   &  -&  -   &  0.015 & 2         \\
6913-0026 & 13.230 &  0.694 & -0.102 &  0.778 &  -     & 
-0.010 & -0.014 &  0.002 &  0.042 &   -  & 4      \\
\noalign{\smallskip}
\hline
\noalign{\smallskip}
& & & & &Mean: &0.000 &0.000 &0.000 &0.000 &0.001 &  \\
& & & & &RMS: &0.023 &0.014 &0.017 &0.017 &0.021 &  \\
\noalign{\smallskip}
\hline 
\end{tabular}
\end{table*}
%_______________________________________________________________________

\subsection{$uvby\beta$ transformation}

The choice of an adequate set of $uvby\beta$ standard stars for CCD
photometry is a very critical issue. By one hand, the primary standard
stars of the $uvby$ and H$\beta$ systems (Crawford \& Mander
\cite{craw66}; Crawford \& Barnes \cite{craw70a}; Perry et
al. \cite{perry87}) are all bright enough to saturate the CCD chip, even
with short exposures. By the other hand, most of the observed stars are
reddened B type stars. Manfroid \& Sterken (\cite{manf87}), Delgado \&
Alfaro (\cite{delgado}) and Crawford (\cite{craw94}) have shown that
transformations made only with unreddened stars introduce large
systematic errors when applied to reddened stars,  even if the colour
range of the standards brackets that of the programme stars.  No such
reddened early type stars are included in the primary $uvby\beta$
standard lists.

Our standard list was composed by stars in young open clusters with
$uvby\beta$ photometry published by Crawford et al. (\cite{craw70b}) for
$h$ \& $\chi$ Persei, Canterna et al. (\cite{canterna}) for NGC 1039 and
Crawford et al. (\cite{craw77}) for NGC 6910 and NGC 6913. All these
photometric lists were obtained with the same telescopes, instrumentation
and reduction procedures used to define the standard Crawford \& Barnes
(\cite{craw66}) and Crawford \& Mander (\cite{craw70a}) 
systems, and so the photometric values are in the standard system. As the
$V$ values for $h$ and $\chi$ Persei are not included in the corresponding
$uvby$ photometric list, we have used the values given by Johnson \&
Morgan (\cite{johnson}). The final standard star list is presented in
Table \ref{t3}. For cluster stars we have adopted the numbering system in
the WEBDA database\footnote{http://obswww.unige.ch/webda/} (Mermilliod
\cite{merm}). Note that these numbers may not correspond with the
numbering in the referred to photometric papers.

Individual stars in each list were selected so as to insure as large 
as possible a range in all photometric indices. However, as the main 
purpose of this paper is to study the clusters upper main sequence, and
in particular the B type spectral range, no late type stars were included in 
the standard star list. A few intrinsically red field stars have been 
observed, and their values have been included in the photometry tables for 
completeness, but it should be noted that photometry for stars redder 
than $(b-y) \sim 0.750$ is likely to be affected by systematic
errors. This 
is also the case for the $\beta$ index of emission-line stars, which reach a
very low value. As no emission line star was included in the standard
list due to their well know variability, there is no way to avoid
extrapolation. 

Transformations were computed from the natural system to the standard
$uvby\beta$ system defined by the standard star list described above. The
obtained transformation equations are the following:

\begin{quote}
$V = -4.793 - 0.023 (b-y)_n + y_n$ \\
$(b-y) = -0.413 + 0.977 (b-y)_n$ \\
$m_1 = 0.063 + 1.044 m_{1,n} - 0.013 (b-y)$ \\
$c_1 = 0.598 + 0.999 c_{1,n} + 0.135 (b-y)$ \\
$\beta = -0.079 + 1.692 \beta_n$ \\  
\end{quote}

were subscript 'n' refers to the natural system. All scale
coefficients in the $uvby$ transformation are close to unit, while the
colour terms are small, indicating a good conformity between the
instrumental and standard photometric systems. This is not the case for
the H$\beta$ transformation, where the scale coefficient is 
larger. This may be due to the wide filter used being significantly
narrower than the standard one defined by Crawford \& Mander
(\cite{craw66}). 

A measure of the photometric accuracy is the standard deviation of the 
mean catalogue minus transformed values for the standard stars. These 
values are presented in Table \ref{t3}, bottom line.
\subsection{Coordinates}
Although precise astrometry is not among the scopes of this paper, we have
transformed the instrumental pixel coordinates into equatorial
coordinates, in order to facilitate the identification of all observed
stars, and their cross-correlation with other photometric lists.

Astronomical coordinates in the field of $h$ Per where derived from 13
stars with positions in the PPM Catalogue (Roeser \& Bastian
\cite{roeser}). In the field of $\chi$ Per there are only 7 stars with
positions in the PPM Catalogue, which are grouped in the eastern part. We
complemented this list with four more stars with positions given by Abad
\& Garc\'\i a (\cite{abad}). Transformation equations were computed by
means of the {\em Starlink} program ASTROM (Wallace \cite{wallace}).
The final astrometric accuracy, measured as the RMS of the mean catalogue
minus transformed values for the stars used in the transformation, is 
within 0\farcs 4.

\subsection{The data}

Equatorial coordinates and mean photometric magnitudes, colours and
indices for stars in the $h$ and $\chi$ Persei nuclei are presented in
Tables 6 and 7. Only frames obtained at airmasses lower than 1.5 have
been considered to compute the final photometric values. For this reason,
small differences up to a few millimagnitudes may appear between values in
Table \ref{t3} and Tables 6 and 7 for the stars in common. 

In the photometric tables we have also adopted the cluster star numbering
system from the WEBDA database. For cluster numbers lower than 3000, WEBDA 
numbers are coincident with Oosterhoff (\cite{oost}) numbers, which we 
will refer to as 'Oo' hereinafter. A few observed stars have not entry in
WEBDA. We have introduced new numbering for them, starting with 7000 in
$h$ Per and with 8000 in $\chi$ Per. 

%________________________________________________________________________
\begin{figure}
\resizebox{\hsize}{!}{\includegraphics{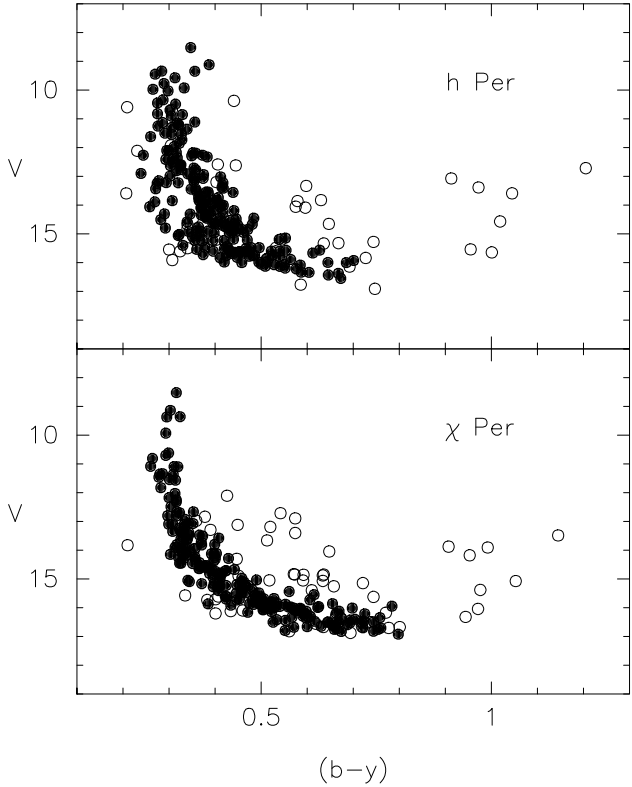}}
\caption{Color-magnitude diagram for $h$ and $\chi$ Persei. Filled
and open circles denote stars considered as cluster members and nonmembers 
respectively} 
\label{f1}
\end{figure}
%_______________________________________________________________________

%________________________________________________________________________
\begin{figure}
\resizebox{\hsize}{!}{\includegraphics{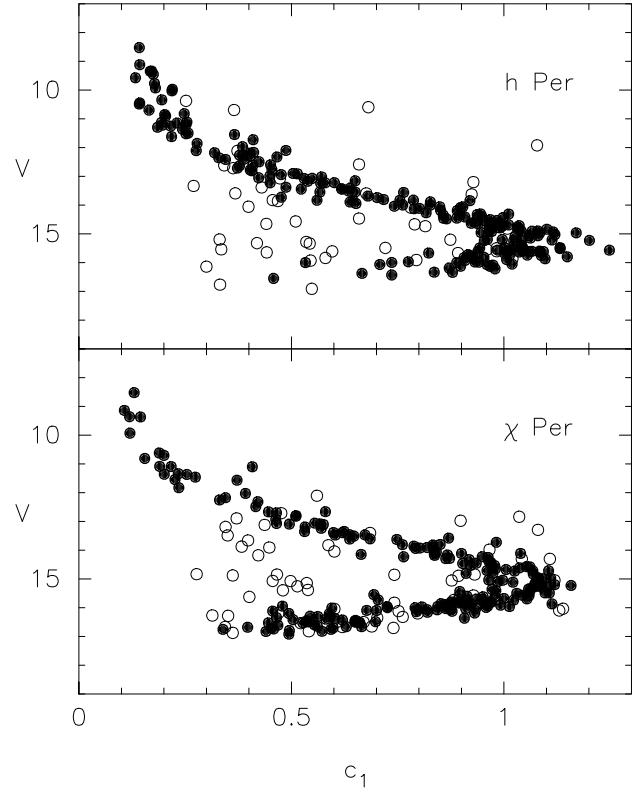}}
\caption{$V-c_1$ photometric diagrams. Symbols as in Fig. \ref{f1}}
\label{f2}
\end{figure}
%_______________________________________________________________________

\section{Reddening, intrinsic colours and distance.}

Colour-magnitude diagrams of all observed stars in both cluster
regions are presented in Fig. \ref{f1},  and photometric
$V-c_1$ diagrams in Fig. \ref{f2}. To obtain the intrinsic colours
we have first classified the stars as belonging to the early (earlier than
A0), intermediate (A0-A3) and late (A3 onwards) groups defined by
Str\"omgren (\cite{strom}). The classification was performed by means of
the algorithm described by Figueras et al. (\cite{figueras}). Cluster
membership were assigned from the position of each star in the different
$uvby\beta$ photometric diagrams. Stars considered as members were marked
in the last column of Tables 6 and 7.

Reddening values and intrinsic colours and indices were obtained for stars 
in the early group by means of the procedure described by Crawford
(\cite{craw78}). We have used the standard $(b-y)_0 - c_0$ relation given
in Table X of Perry et al. (\cite{perry87}), and the following
reddening relations:

\begin{quote}
$A_V = 4.3 E(b-y)$ \\
$E(c_1) = 0.2 E(b-y)$ \\
$E(m_1) = -0.32 E(b-y)$ \\
\end{quote}

Known Be stars have not been included in the computation of the
interstellar reddening, as they present an additional reddening
contribution of circumstellar origin.

\begin{figure}
\resizebox{\hsize}{!}{\includegraphics{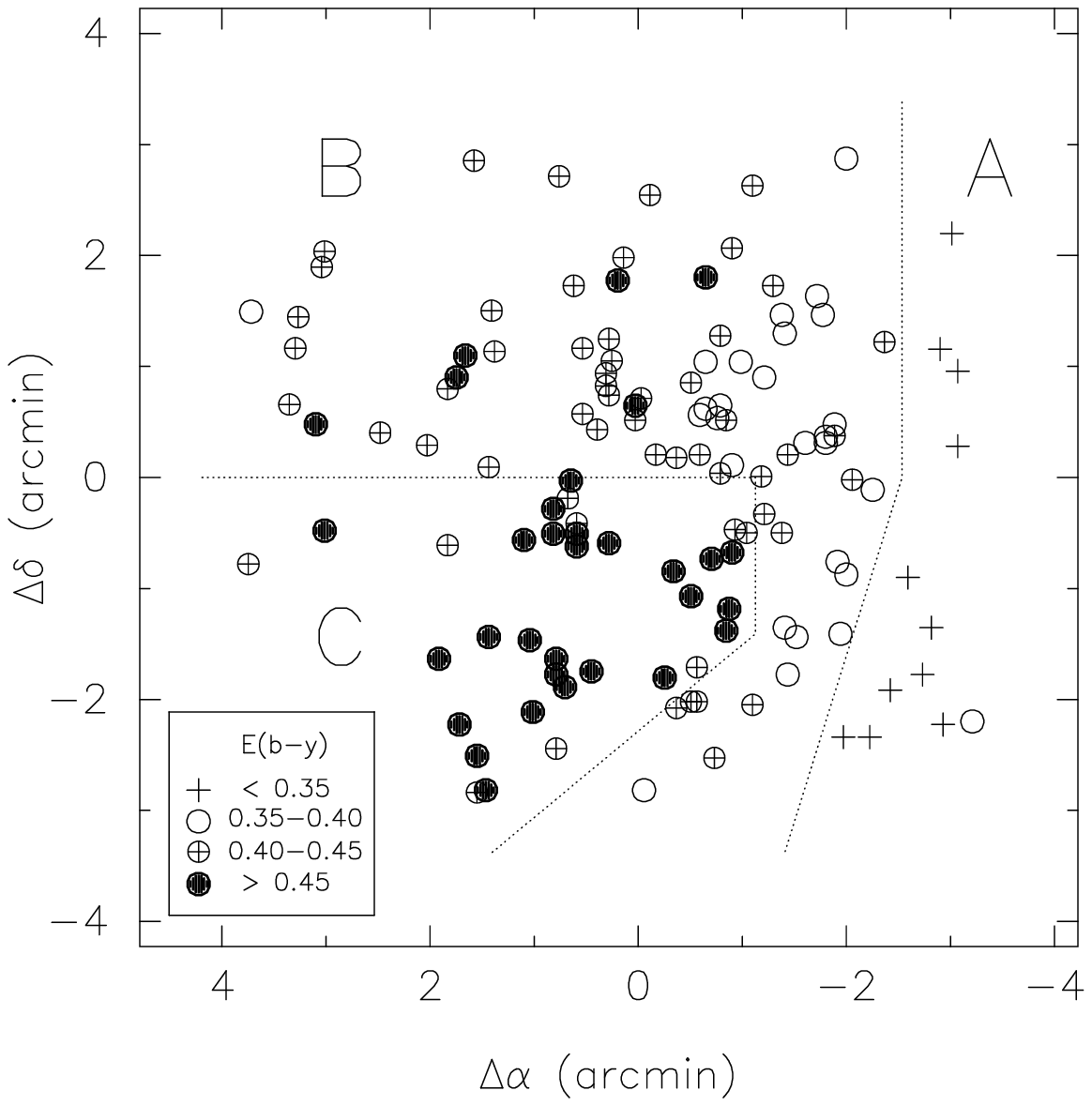}}
\caption{Reddening spatial distribution for B stars in the  $h$ Per
nucleus. Positions are relative to star Oo 1057 
(02$^{\rm h}$19$^{\rm m}$04\fs 45, +57\degr 08\arcmin 07\farcs 8,
J2000)}
\label{f3}
\end{figure}

For $h$ Per we obtained a mean reddening value of $E(b-y) = 0.420 \pm
0.045$, from 127 stars. The large value of the standard deviation
indicates the presence of variable reddening across the cluster
nucleus. In Fig. \ref{f3} we have represented the reddening values for
individual stars as a function of their position. It is apparent a trend
of increasing reddening from west to east, and a heavily obscured region
in the south-eastern part of the cluster center. In Fig. \ref{f3} we have
divided the cluster nucleus area in three regions of different 
reddening. Mean reddening values in these regions are presented in Table
\ref{t6}. 

The large absorption in the south-east part has been previously pointed
out by Waelkens et al. (\cite{waelkens}) from photoelectric photometry
in the Geneva system, but surprisingly has not been accounted for in 
recent CCD studies. On the other hand, the low reddening value in the 
western part was already noted by Fabregat et al. (\cite{fabregat96}). 
They considered the Be stars \object{Oo 146}, \object{245} and
\object{566} as probable nonmembers, 
due to their reddening values ($E(b-y) =$ 0.313, 0.346 and 0.308 
respectively), which are much lower than the cluster average reddening. 
From the present result we can conclude that the low reddening values are 
consistent with the position of these stars in the low absorption region 
at the western part of the cluster.

\begin{figure}
\resizebox{\hsize}{!}{\includegraphics{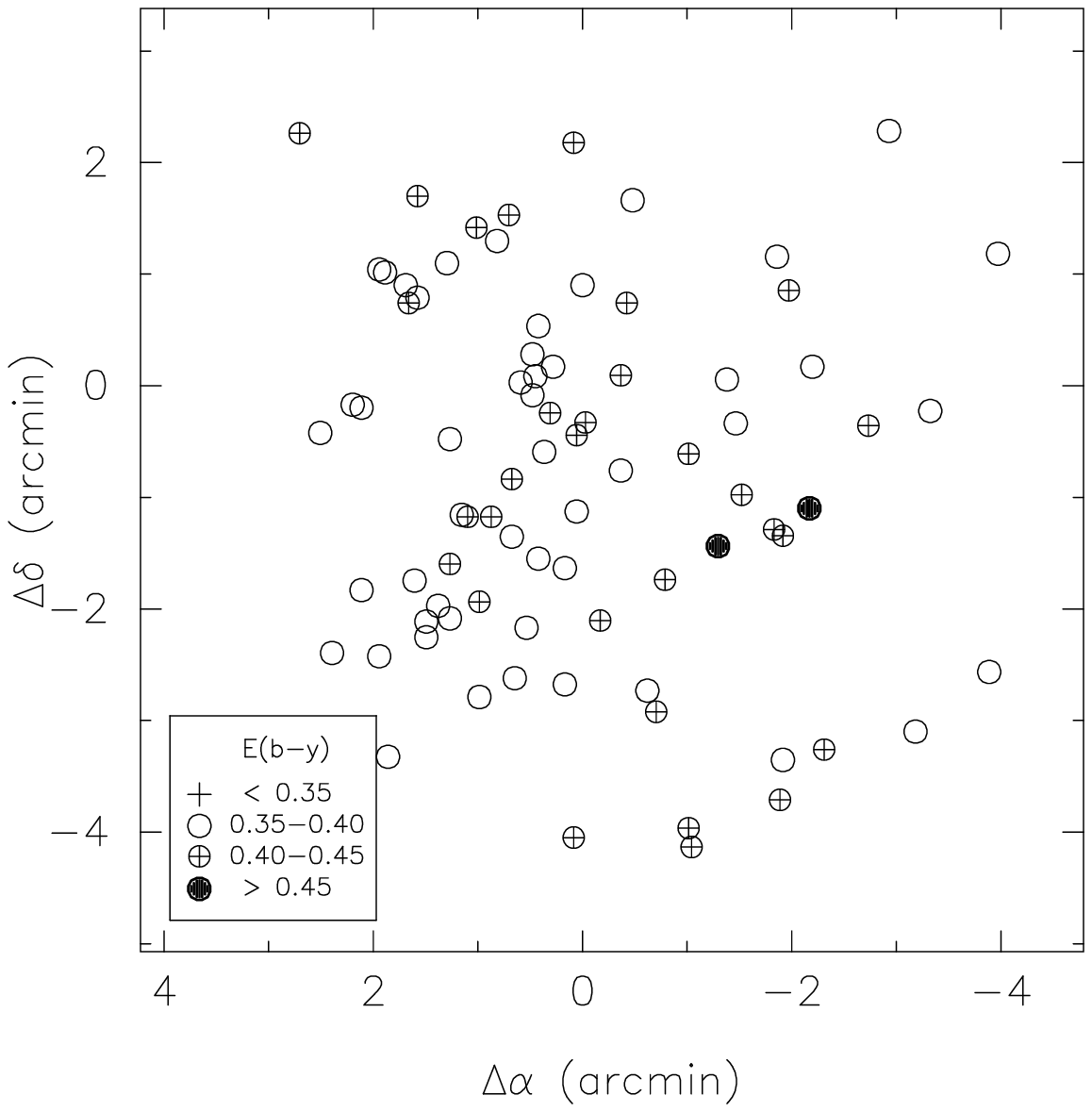}}
\caption{Reddening spatial distribution for B stars in the $\chi$ Per 
nucleus. Positions are relative to star Oo 2227 
(02$^{\rm h}$22$^{\rm m}$00\fs 59, +57\degr 08\arcmin 41\farcs 9,
J2000)}
\label{f4}
\end{figure}

The mean value of the reddening for $\chi$ Per is $E(b-y) = 0.398 \pm
0.025$, from 82 stars. In Fig. \ref{f4} we have also represented the
individual reddening values as a function of the stars position. In this
case no trend is present. The standard deviation of the mean reddening
value is only slightly larger than the deviation of the reddening 
computation method (namely 0.018, cf. Perry \& Johnston \cite{perry82}; 
Franco \cite{franco}), and we conclude that the interstellar absorption
is constant towards the $\chi$ Per nucleus, within the accuracy of our
photometry.

The different reddening characteristics towards both cluster nuclei are 
apparent in the photometric diagrams presented in Figs. \ref{f1} and 
\ref{f2}. In the $V-(b-y)$ plane, the $h$ Per sequence is much broader 
than the $\chi$ Per one, due to the strongly variable reddening in the 
former. Assuming that both clusters are at the same distance (see below), 
the higher mean reddening in $h$ Per causes that its observed sequence 
ends at earlier spectral types. This is apparent in Fig. \ref{f1}, but 
much more remarkable in the $V-c_1$ plane in Fig. \ref{f2}: in $\chi$ Per 
the lower branch of the cluster sequence is almost complete until $c_1 =
0.45$ (spectral type around F5), while in $h$ Per it ends at $c_1 = 0.85$
(around A7).  

%________________________________________________________________________
%_______________________________________________________________________
%________________________________________________________________________
%_______________________________________________________________________

%__________________________________________________________________________ 
\begin{table}
\centering
\caption[]{Mean reddening values for the three regions of the $h$ Per
nucleus defined in Fig. \ref{f3}}
\label{t6}
\begin{tabular}{ccc}
\hline 
\noalign{\smallskip}
Region & $E(b-y)$ & stars  \\
\noalign{\smallskip}
\hline
\noalign{\smallskip}
A  & $0.328 \pm 0.022$ & 12 \\
B  & $0.414 \pm 0.026$ & 81 \\
C  & $0.465 \pm 0.024$ & 34 \\
\noalign{\smallskip}
\hline 
\end{tabular}
\end{table}  
%_____________________________________________________________________________

\begin{figure}
\resizebox{\hsize}{!}{\includegraphics{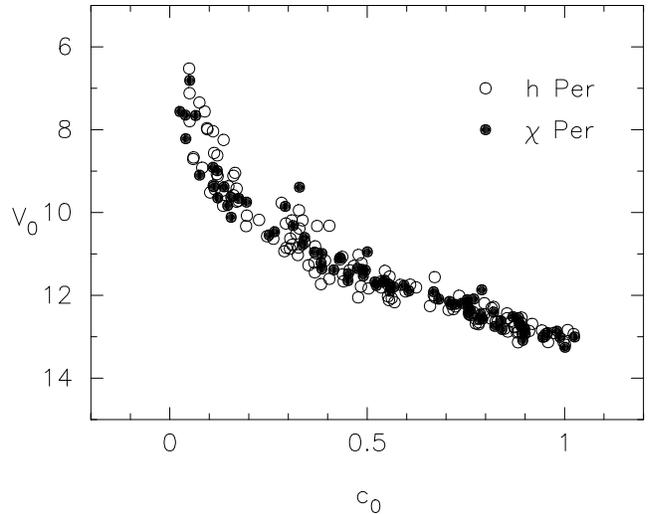}}
\caption{Intrinsic $V_0 - c_0$ diagram for B stars in $h$ and $\chi$ Per
nuclei}
\label{f5}
\end{figure}

In Fig. \ref{f5} we present the intrinsic $V_0 - c_0$ diagram for B stars
in both cluster nuclei. $\chi$ Per stars have been dereddened by using the
mean cluster value of $E(b-y) = 0.398$. For $h$ Per, each star has been
dereddened on the basis of its position within Fig. \ref{f3}, and using
the reddening values in Table \ref{t6}. The loci of both sequences are
undistingishable, indicating that both clusters are at the same
distance, and have also the same age, as we will discuss in the next
section. The sequence of $h$ Per is somewhat broader, most likely due to
the variable reddening. 

\begin{figure}
\resizebox{\hsize}{!}{\includegraphics{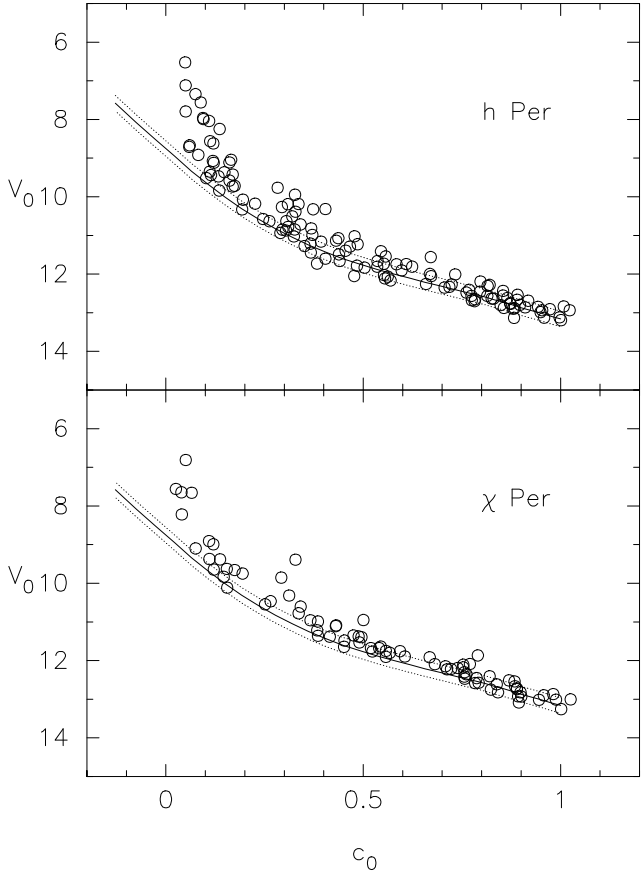}}
\caption{ZAMS fitting to the $h$ and $\chi$ Per B star sequences. The
solid line represents the ZAMS at a distance modulus of 11.7 mag. Dashed
lines are for distance moduli of 11.5 and 11.9 respectively.}
\label{f6}
\end{figure}

To obtain the distance we have fitted to the $V_0 - c_0$ diagram of each
cluster  the ZAMS as presented in Table X of Perry et al. 
(\cite{perry87}). Results are shown in Fig. \ref{f6}. We found the best 
fit at a distance modulus of 11.7 mag. To estimate the error of this 
determination we have also represented in Fig. \ref{f6}, as dotted lines, 
the ZAMS shifted by distance moduli of 11.5 and 11.9 respectively. We 
find that the 11.7 value produces a distinctly better fit that the two 
latter ones, and hence we give the value of $11.7 \pm 0.1$ mag. as the 
distance modulus of $h$  and $\chi$ Persei. This value is in good agreement 
with the recent determinations based on CCD photometry by Keller et al. 
(\cite{keller}) and Marco \& Bernabeu (\cite{marco}). 

%________________________________________________________________________
%_______________________________________________________________________

\begin{figure}
\resizebox{\hsize}{!}{\includegraphics{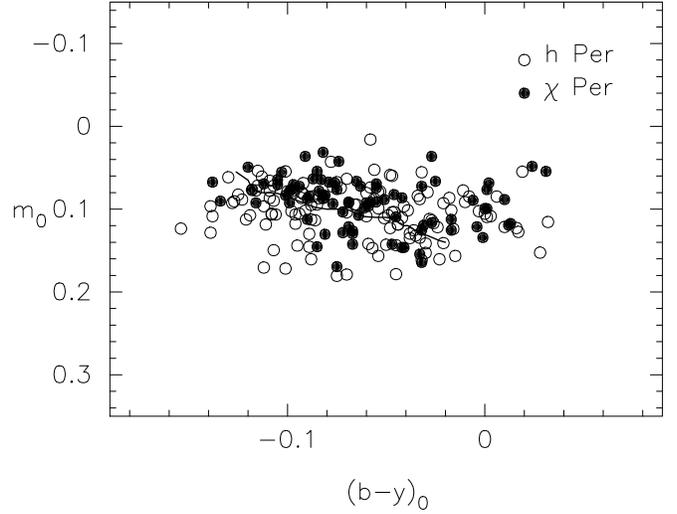}}
\caption{$m_0 - (b-y)_0$ diagram for B type stars in $h$ and $\chi$
Per nuclei, with the mean relation for nearby stars.}
\label{f7}
\end{figure}

\begin{figure}
\resizebox{\hsize}{!}{\includegraphics{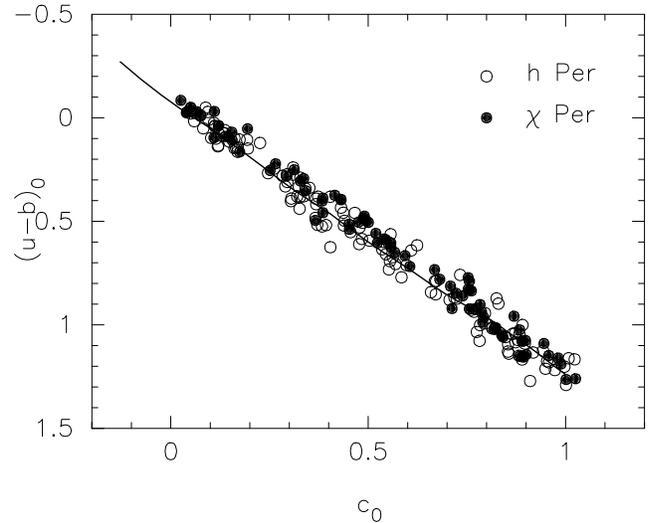}}
\caption{$(u-b)_0 - c_0$ diagram for B type stars in $h$ and $\chi$
Per nuclei, with the mean relation for nearby stars.}
\label{f8}
\end{figure}

Finally, an indication of the validity of the dereddening procedure and
the conformity between our photometry and the standard system can be
obtained by comparing the resulting $m_0$ and $(u-b)_0$ indices with those
of nearby, unreddened stars as given by Perry et al. (\cite{perry87}). The
results are shown in Figs. \ref{f7} and \ref{f8}. As it can be seen, the
obtained indices are coincident with the mean intrinsic relations, and
hence we can conclude that our photometry is in the standard system, and
free of systematic effects. This is in turn a proof of the validity of our
reduction methods and the adequacy of the standard star selection.

%________________________________________________________________________
%_______________________________________________________________________

\section{Cluster ages}

Determination of the cluster ages will be done by isochrone fitting to
the upper main sequence. In the $uvby$ system and for stars in the early
group, the $(b-y)_0$ colour and the $c_0$ index are temperature
indicators, and hence both $V_0 - (b-y)_0$ and $V_0 - c_0$ planes are
observational HR diagrams. Following the discussion in Fabregat \&
Torrej\'on (\cite{fabregat00}) we consider more precise and reliable the
isochrone fitting to the $V_0 - c_0$ diagram, for the following
reasons: i./ the range of variation of the $c_0$ index along the B-type
sequence is more than ten times larger than the range of variation of
$(b-y)_0$, providing much better discrimination between isochrones of
similar ages; ii./ the $c_0$ index is less affected, by a factor of 5, by
interstellar reddening; iii./ the $V_0 - c_0$ plane allows an efficient
segregation of emission line stars.

%________________________________________________________________________
%_______________________________________________________________________
%________________________________________________________________________
%_______________________________________________________________________

\begin{figure}
\resizebox{\hsize}{!}{\includegraphics{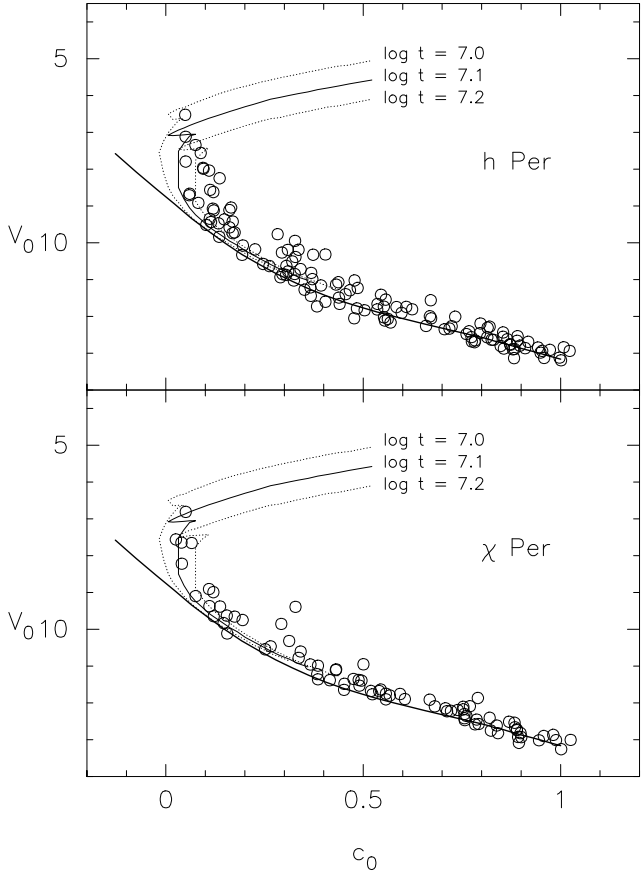}}
\caption{Isochrone fitting to the $h$ and $\chi$ Per B star
sequences, in the $V_0 - c_0$ plane.} 
\label{f9}
\end{figure}

In Fig. \ref{f9} we present the $V_0 - c_0$ sequences of $h$ and $\chi$
Per, together with isochrones with ages of $\log t =$ 7.0, 7.1 and 7.2 
years. The isochrones have been computed with the evolutionary models of
Schaller at al. (\cite{schaller}), and transformed  to the observational
plane by means of the relations obtained  by Torrej\'on (\cite{torrejon}).   
The best fitting in both clusters is obtained for $\log t =$ 7.1, with
most of the stars at the turnoff point lying between the $\log t =$ 7.0
and 7.2 isochrones. Hence we propose $\log t = 7.10\pm0.05$ as the common
age of both clusters, in good agreement with the recent result of Keller
et al. (\cite{keller}).
 
%________________________________________________________________________
%_______________________________________________________________________

\section{Discussion}

In a recent paper Marco \& Bernabeu (\cite{marco}, hereinafter referred to
as MB01) claimed that $h$ Per is significantly older than $\chi$ Per, and
go further to propose three epochs of star formation within the double
cluster at ages of log $t$ = 7.0, 7.15 and 7.3 years. This is in clear
disagreement with our results in the previous section. Their study is also
based on CCD $uvby$ photometry, and they use the same methods and
techniques that we do in the present paper. Hence, it is worth to compare
both studies in order to ascertain the reasons of the diverging results.
When comparing both sets of data, it should be noted that: i./ our
standard photometry is at least twice as accurate as the MB01 one
(cf. Table 9 in MB01 and Table \ref{t3} in this paper), and ii./our CCD 
frames cover a larger area than MB01 ones, and hence our photometric 
sequences are more populated.

\begin{figure}
\resizebox{\hsize}{!}{\includegraphics{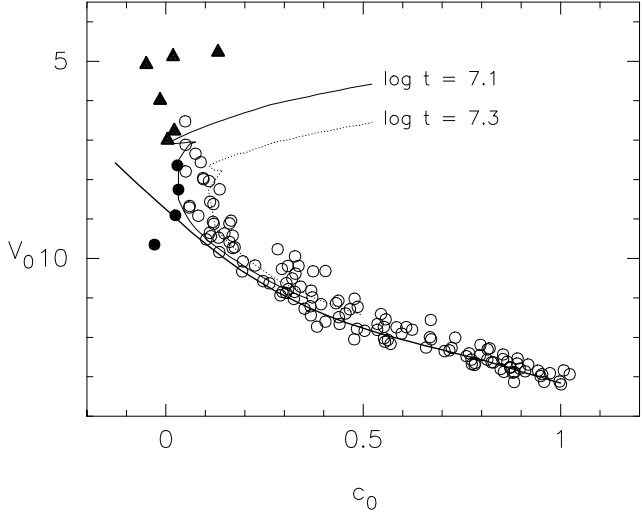}}
\caption{$V_0 - c_0$ diagram for B type stars in $h$ Per. Open circles
are main sequence stars, filled circles Be stars and triangles supergiant
stars.} 
\label{f10}
\end{figure}

MB01 find an age of log $t$ = 7.10 -- 7.15 years for $\chi$ Per, in
agreement with our value of log $t$ = 7.10 obtained in the previous
section. For $h$ Per they obtain a significantly older age. In their
Fig. 10 they show the log $t$ = 7.3 isochrone fitting a clump of stars at
about $c_0 = 0.12$, which they consider as the main sequence turnoff. This
clump is also present in our data, as can be seen in Fig. \ref{f10} where
we present the $V_0 - c_0$ plane and the the log $t$ = 7.1 and 7.3
isochrones. However, the distribution of stars in our Fig. \ref{f10} near
the turnoff point is different. The clump at $c_0 = 0.12$ is less
conspicuous, it is only formed for at most six stars, and significantly
leftwards to the fitting log $t$ = 7.3 isochrone there are at least eight
main sequence cluster stars, which should have a younger age. We consider
the latter as defining the actual cluster main sequence turnoff, which is
well fitted by the log $t$ = 7.1 isochrone. 

The possibility of two distinct populations with log $t$ = 7.1 and 7.3
years seems very unlikely. The occurrence of stars slightly above the main
sequence can be justified by several reasons, including binarity, high
rotational velocity, or photometric errors. Other small clump above
the main sequence is apparent at $c_0 = 0.3$, and it is also most likely
caused by the above referred to reasons. 

MB01 also propose the presence of a population younger than log $t$ = 7.0
in the $h$ Per region. This claim is based on the position in the $V_0 -
c_0$ plane of six supergiant B stars observed by Crawford et
al. (\cite{craw70b}). In Fig. \ref{f10} we have also represented these
stars (filled triangles), which clearly fall at the left of the log $t$ =
7.1 isochrone. 

But this argument is wrong. The $c_1$ index of the $uvby$ system is
defined to be a measure of the
Balmer discontinuity depth, which for OB stars is related to the
effective temperature. For this reason, the $V_0 - c_0$ photometric plane  
can be considered as an observational Hertzsprung-Russell
diagram. However, this is only true for stars of luminosity classes III
to V, and not for supergiants. Crawford et al. (\cite{craw70b}) 
already note that Ia type supergiants do not follow the mean 
$(b-y)_0 - c_0$ calibration. Schild \& Chaffee (\cite{schild75}) 
show that the Balmer continuum of early type supergiants is affected by
an emission or absorption contribution of circumstellar origin, and hence
the Balmer discontinuity depth is not correlated with spectral types or
effective temperature. Kilkenny \& Whittet (\cite{kilkenny}) present 
$(b-y)_0 - c_0$ intrinsic relations for early type supergiants. There 
is a  different relation for each luminosity class (Ia, Iab, Ib and II), 
and all of them are in turn different to the mean relation for III-V
classes. If we assume that $(b-y)_0$ is a good temperature indicator for all 
luminosity classes (see below), this implies that the $c_0$ index cannot 
define a unique temperature scale valid for all luminosities.

The case is much the same as for classical Be stars. Be stars also have
continuum emission or absorption of circumstellar origin at the Balmer
discontinuity, and hence their $c_0$ indices are anomalous (Fabregat et
al. \cite{fabregat96}, and references therein). For comparison, in
Fig. \ref{f10} we have represented the four Be stars in the cluster
nucleus (filled circles). They also occupy anomalous positions at the left
of the isochrones, and even below the ZAMS. Neither supergiants nor Be
stars should be used in a $V_0 - c_0$ diagram when it is
intended to be an observational HR diagram for isochrone fitting
purposes, due that their $c_0$ index is not related with effective
temperature.   

%________________________________________________________________________
%_______________________________________________________________________

%________________________________________________________________________
\begin{figure}
\resizebox{\hsize}{!}{\includegraphics{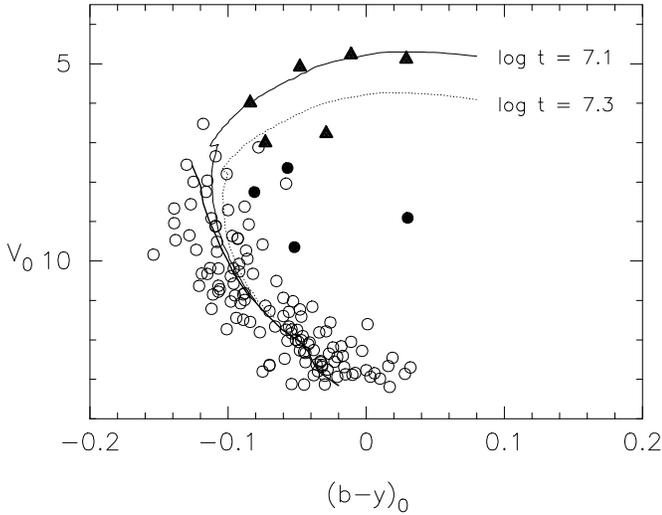}}
\caption{$V_0 - (b-y)_0$ plane for B type stars in $h$ Per. Symbols as in 
Fig. \ref{f10}.}
\label{f11}
\end{figure}
%_______________________________________________________________________

In order to study the age of the supergiant star population in the $h$ Per
area, in Fig. \ref{f11} we have represented the $V_0 - (b-y)_0$
photometric diagram. $(b-y)$ is a measure of the Paschen continuum slope,
which is correlated with effective temperature for stars of all luminosity
classes. Be stars, however, also deviates due to additional reddening of
circumstellar origin (eg. Fabregat et al. \cite{fabregat96}). Two
supergiant stars, \object{Oo 1057} and \object{1162}, lie in the cluster
nucleus, and have been dereddened as a function of their position in
Fig. \ref{f3}. The other four, \object{Oo 3}, \object{16}, \object{612} 
and \object{662}, are placed at the west of the cluster
nucleus. Following the discussion in Sect. 3, we have used the value of
$E(b-y) = 0.328$ in Table \ref{t6} to obtain their intrinsic colours. In 
order to check the reliability of this value, we have calculated the
individual reddening for all supergiant stars using the standard relations
given by Kilkenny \& Whittet (\cite{kilkenny}). Results are presented in
Table \ref{t7}. All reddening values for the western stars are compatible 
with the mean reddening for the west region given in Table \ref{t6}.

%__________________________________________________________________________ 
\begin{table}
\centering
\caption[]{Reddening values for blue supergiant stars observed by Crawford
et al. \cite{craw70b} }
\label{t7}
\begin{tabular}{rc}
\hline 
\noalign{\smallskip}
Oo & $E(b-y)$   \\
\noalign{\smallskip}
\hline
\noalign{\smallskip}
3 & 0.306 \\
16 & 0.340 \\
612 & 0.350 \\
662 & 0.357 \\
1057 & 0.411 \\
1162 & 0.473 \\
\noalign{\smallskip}
\hline
\end{tabular}
\end{table}  
%_____________________________________________________________________________

In Fig. 11 it is apparent that most of the supergiant stars lie close to
the log $t$ = 7.1 isochrone, giving further support to this value as the
actual age of $h$ Per. There are no stars significantly leftwards to this
isochrone, indicating that no younger population is present in the cluster
area.

MB01 also presented the $V_0 - (b-y)_0$ plane for $h$ Per (their Fig. 9). 
In their figure most supergiant stars are placed leftwards to the log
$t$ = 7.0 isochrone, and even leftwards to the ZAMS, indicating that these
stars could be no members of the cluster. This is due to a
reddening overcorrection. They used their mean cluster value,
$E(b-y) = 0.44$, to calculate the intrinsic colours. The deviating 
supergiants are placed at the west of the cluster nucleus, where we have
demonstrated that the reddening is lower by more than a tenth of
magnitude. With the proper reddening values supergiants are placed close
to the cluster isochrone.

We can conclude that there are no different epochs of star formation
within the $h$ and $\chi$ Persei clusters area, and that both clusters
and the surrounding field stars share the same age of $\log t =
7.10\pm0.05$ years, as determined in the previous sections.  We propose
that the wrong results obtained by MB01 are due to a number of factors,
including the low accuracy of their photometry, the neglecting of the
strongly variable reddening across the $h$ Per cluster, the incorrect use
of supergiant stars in the $V_0 - c_0$ plane for isochrone analysis, and
to a large extent the overinterpretation of the data at their disposal. 

As discussed in Sect. 4, and in more detail in Fabregat \& Torrej\'on
(\cite{fabregat00}), the isochrone fitting in the $V_0 - c_0$ plane is an
excellent tool for accurate age determination in young open clusters. With
accurate photometry, the main sequence turnoff is very well defined,
allowing a precise discrimination between isochrones of slightly different
ages. But its applicability is limited to the range of spectral types and
luminosity classes for which the $c_0$ index is a good temperature
indicator. This is restricted to OB stars of luminosity classes III to V,
and beyond this limits its use can conduct to misleading results.

\section{Conclusions}
 
We have presented CCD $uvby\beta$ photometry for stars in the nuclei of
the young open clusters $h$ and $\chi$ Per. We have shown that our
photometry is free of systematic effects and well tied to the standard
$uvby$ system.

We have obtained the cluster
astrophysical parameters from the analysis of the B type star range.
The reddening is highly variable through the $h$ Per nucleus, increasing
from west to east. Its value ranges from $E(b-y) = 0.328\pm0.022$ in the
western part to $E(b-y) = 0.465\pm0.024$ in the south-east. Towards  
$\chi$ Per the reddening is fairly constant, with $E(b-y) = 
0.398\pm0.025$. Both clusters share a common distance modulus of
11.7$\pm$0.1 mag., and an age of $\log t = 7.10\pm0.05$ years.

\begin{acknowledgements}
We would like to thank Dr. J. Manfroid for providing us with his RANBO2
code to compute atmospheric extinction and photometric natural system. We
are grateful to the Observatorio Astron\'omico National for the allocation
of observing time in the 1.5m. telescope, and for support during 
observations. 
This research has made use of the WEBDA database, developed and maintained
by J.C. Mermilliod, the SIMBAD database, operated at CDS, Strasbourg,
France, and the NASA's Astrophysics Data System Abstract Service.
The authors acknowledge the data analysis facilities provided by the
IRAF data reduction and analysis system, and by the Starlink Project which
is run by CCLRC on behalf of PPARC.
This work has been partially supported by the {\it Plan Nacional de
Investigaci\'on Cient\'\i fica, Desarrollo e Innovaci\'on Tecnol\'ogica
del Ministerio de Ciencia y Tecnolog\'\i a} and FEDER, through contract
AYA2000-1581-C02-01.
JF acknowledges grants from the {\it Conselleria de Cultura i Educaci\'o
de la Generalitat Valenciana} and the {\it Secretar\'\i a de Estado de
Educaci\'on y Universidades} of the Spanish Governement. 
  
\end{acknowledgements}


\begin{thebibliography}{}
\bibitem[1995]{abad}
 Abad, C., Garc\'\i a, L. 1995, Rev. Mex. Astron. Astrophys. 31, 15

%\bibitem[1975]{baliunas}
%  Baliunas S.L., Ciccione M.A., Guinan E.F. 1975, PASP 87, 969

%\bibitem[1994]{balona}
%  Balona L.A. 1994, MNRAS 267, 1060

%\bibitem[1994]{balonakoen}
%  Balona L.A., Koen C. 1994, MNRAS 267, 1071

%\bibitem[1984]{balonashob}
%  Balona L.A., Shobbrook R.R. 1984, MNRAS 211, 375

%\bibitem[1994]{bertelli}
%  Bertelli G., Bressan A., Chiosi C. et al. 1994, A\&AS 106, 275

\bibitem[1979]{canterna}
  Canterna, R., Perry, C.L., Crawford, D.L. 1979, PASP 91, 263

%\bibitem[1974]{claria}
%  Claria J.J. 1974, Elementos de fotometr\'ia estelar, Instituto
%  Venezolano de Astronom\'ia

%\bibitem[1975]{craw75}
%  Crawford D.L. 1975, AJ 80, 955

\bibitem[1978]{craw78}
  Crawford, D.L. 1978, AJ 83,48

%\bibitem[1979]{craw79}
%  Crawford D.L. 1979, AJ 84, 1858

\bibitem[1994]{craw94}
  Crawford, D.L. 1994, PASP 106, 397

\bibitem[1966]{craw66}
  Crawford, D.L., Mander, J. 1966, AJ 71, 144

\bibitem[1970]{craw70a}
  Crawford, D.L., Barnes, J.V. 1970, AJ 75, 978

\bibitem[1970]{craw70b}
  Crawford, D.L, Glaspey, J.W., Perry, C.L. 1970, AJ 75, 822

\bibitem[1977]{craw77}
  Crawford, D.L., Barnes, J.V., Hill G. 1977, AJ 82, 606

\bibitem[1989]{delgado}
  Delgado, A.J., Alfaro, E.J. 1989, A\&A 219, 121

\bibitem[2000]{fabregat00}
  Fabregat, J., Torrej\'on, J.M., 2000, A\&A 357, 451

\bibitem[1996]{fabregat96}
  Fabregat, J., Torrej\'on, J.M., Reig, P. et al. 1996, A\&AS 119, 271

\bibitem[1991]{figueras}
  Figueras, J., Torra, J., Jordi, C. 1991, A\&AS 87, 319

%\bibitem[1993]{fitzsim1}
%  Fitzsimmons A. 1993, A\&AS 99,15

\bibitem[1989]{franco}
  Franco, G.A.P. 1989, A\&AS 78, 105

%\bibitem[1976]{gronbech}
%  Gr$\o$nbech B., Olsaen E.H., Str\"omgrem B. 1976, A\&AS 26, 155

%\bibitem[1992]{iglesias}
%  Iglesias C.A., Rogers F.j., Wilson B.G. 1992, ApJ 397, 717

\bibitem[1955]{johnson}
  Johnson, H.L., Morgan, W.W. 1955, ApJ 122, 429

%\bibitem[1997]{jordi}
%  Jordi, C. Masana, E., Figueras, F., Torra, J. 1997, A\&AS 123, 83

\bibitem[2001]{keller}
  Keller, S.C., Grebel, E.K., Miller, G.J., Yoss, K.M. 2001, AJ 122, 248

\bibitem[1985]{kilkenny}
  Kilkenny, D., Whittet, D.C.B. 1985, MNRAS 216, 127

\bibitem[1993]{manf93}
  Manfroid, J. 1993, A\&A 271, 714

\bibitem[1987]{manf87}
  Manfroid, J., Sterken, C. 1987, A\&AS 71, 539

\bibitem[2001]{marco}
  Marco, A., Bernabeu, G. 2001, A\&A 372, 477

\bibitem[1999]{merm}
  Mermilliod, J.C. 1999, in Very Low-Mass Stars and Brown Dwarfs in
Stellar Clusters and Associations, Cambridge Univ. Press, eds. Rebolo,
R. and Zapatero-Osorio, R.M.

%\bibitem[1993]{meynet}
%  Meynet G., Mermilliod J.C., Maeder A., 1993, A\&AS 98, 477 

%\bibitem[1983]{muminov}
% Muminov M. 1983, BICDS 24 

%\bibitem[1984]{olsen}
%  Olsen E.H. 1984, A\&AS 57, 443

\bibitem[1937]{oost}
  Oosterhoff, P.T. 1937, Ann. Sterrewatch Leiden 17, 1

\bibitem[1982]{perry82} 
  Perry, C.L., Johnston, L. 1982, ApJS 50, 451 

\bibitem[1987]{perry87} 
  Perry, C.L., Olsen, E.H., Crawford, D.L. 1987, PASP 99, 1184 

\bibitem[1988]{roeser}
  Roeser, S., Bastian, U. 1988, A\&AS 74, 449 

\bibitem[1992]{schaller}
 Schaller, G., Schaerer, D., Meynet, G., Maeder, A. 1992, A\&A 96, 269

%\bibitem[1967]{schild67}
%  Schild R.E. 1967, ApJ 148, 449

\bibitem[1975]{schild75}
  Schild, R.E., Chaffee, F.H. 1975, ApJ 196, 503

\bibitem[1987]{stetson}
  Stetson, P.R. 1987, PASP 99, 191

%\bibitem[1992]{stetson1}
%  Stetson P.R. 1992, User's Manual for DAOPHOT II

\bibitem[1966]{strom}
  Str\"omgrem, B., 1966, ARA\&A 4, 433

\bibitem[1997]{torrejon}
  Torrej\'on, J.M. 1997, PhD Thesis, University of Valencia

\bibitem[1990]{waelkens}
  Waelkens, C., Lampens, P., Heynderickx, D. et al. 1990, A\&AS 83, 11

\bibitem[1998]{wallace}
  Wallace, P.T. 1998, Starlink User Note 5.17, Rutherford Appleton
  Laboratory

\end{thebibliography}
 \end{document}